\documentclass[aps,prl,twocolumn,showpacs,preprintnumbers,amsmath,amssymb,superscriptaddress,floatfix]{revtex4}

\usepackage{graphicx}
\usepackage{amsmath}

\renewcommand{\Re}{\mathrm{Re}}
\renewcommand{\Im}{\mathrm{Im}}

\newcommand{\Ep}{\mathcal{E}_p}
\newcommand{\Ec}{\mathcal{E}_c}
\newcommand{\Eo}{\mathcal{E}_0}

\begin{document}

\title{Ultraslow Propagation of Matched Pulses by Four-Wave Mixing in
an Atomic Vapor}

\author{V. Boyer}
\affiliation{Atomic Physics Division, MS 8424, National Institute of
Standards and Technology, Gaithersburg, Maryland 20899-8424, USA}
\author{C. F. McCormick}
\affiliation{Atomic Physics Division, MS 8424, National Institute of
Standards and Technology, Gaithersburg, Maryland 20899-8424, USA}
\author{E. Arimondo}
\affiliation{Atomic Physics Division, MS 8424, National Institute of
Standards and Technology, Gaithersburg, Maryland 20899-8424, USA}
\affiliation{Dipartimento di Fisica Enrico Fermi, Universit\`a di Pisa,
Largo Bruno Pontecorvo 3,
I-56127 Pisa, Italy}
\author{P. D. Lett}
\affiliation{Atomic Physics Division, MS 8424, National Institute of
Standards and Technology, Gaithersburg, Maryland 20899-8424, USA}

\date{\today}

\begin{abstract}

We have observed the ultraslow propagation of matched pulses in
nondegenerate four-wave mixing in a hot atomic vapor.  Probe pulses as
short as 70~ns can be delayed by a tunable time of up to 40~ns with
little broadening or distortion. During the propagation, a probe pulse 
is amplified
and generates a conjugate pulse which is faster and separates from the probe
pulse before getting locked to it at a fixed delay.  The precise
timing of this process allows us to determine the key coefficients of the
susceptibility tensor.  The presence of gain in this system makes this 
system very
interesting in the context of all-optical information processing. 

\end{abstract}


\maketitle

Slow group velocities, valuable for all-optical signal processing, are
obtained at a resonance peak of the transmission spectrum of a medium,
and a number of different implementations of this principle have been
demonstrated.  They rely either on a reduction of the absorption, such
as electromagnetically induced transparency (EIT)~\cite{kasapi1995},
coherent population oscillations~\cite{bigelow2003}, and dual
absorption lines~\cite{camacho2006}, or on a gain resonance, like
stimulated Brillouin scattering~\cite{okawachi2005} and stimulated
Raman scattering~\cite{sharping2005}.  To be useful in the context of
all-optical signal processing, an optical delay line should be able to
produce a fractional delay (defined as the ratio of the delay to the
duration of the pulse) larger than unity with only modest absorption
and pulse broadening.  Recent
developments~\cite{okawachi2005,zhang2006,jiang2006} have shown the
benefits of using an amplifying medium to alleviate the absorption and
distortion issues usually associated with slow
light~\cite{matsko2005}.  

We have examined the group velocity reduction effects
due to nondegenerate four-wave mixing (4WM) in hot rubidium vapor, and
have obtained large fractional delays with almost no distortion.  The
presence of gain in this system makes it in principle possible to stack such delay
lines and achieve fractional delays only limited by pulse broadening. 
Another notable feature of the amplification in the 4WM process is the 
generation of a conjugate pulse which is coupled to the probe and which
propagates alongside it, similar to matched pulses in EIT
systems~\cite{harris1997}.  We have studied the interplay between the
4WM coupling of the probe and conjugate, and the Raman coupling
of the probe and pump. This interplay leads to the ultraslow propagation of
matched probe and conjugate pulses and to the enhancement of the 4WM
gain.  Such an enhancement, when applied to cross-phase modulation, is
the key element of recent optical quantum information processing
proposals~\cite{lukin2000b,lukin2001,wang2006}.

\begin{figure}[htb]
    \begin{center}
	\includegraphics[width=\linewidth]{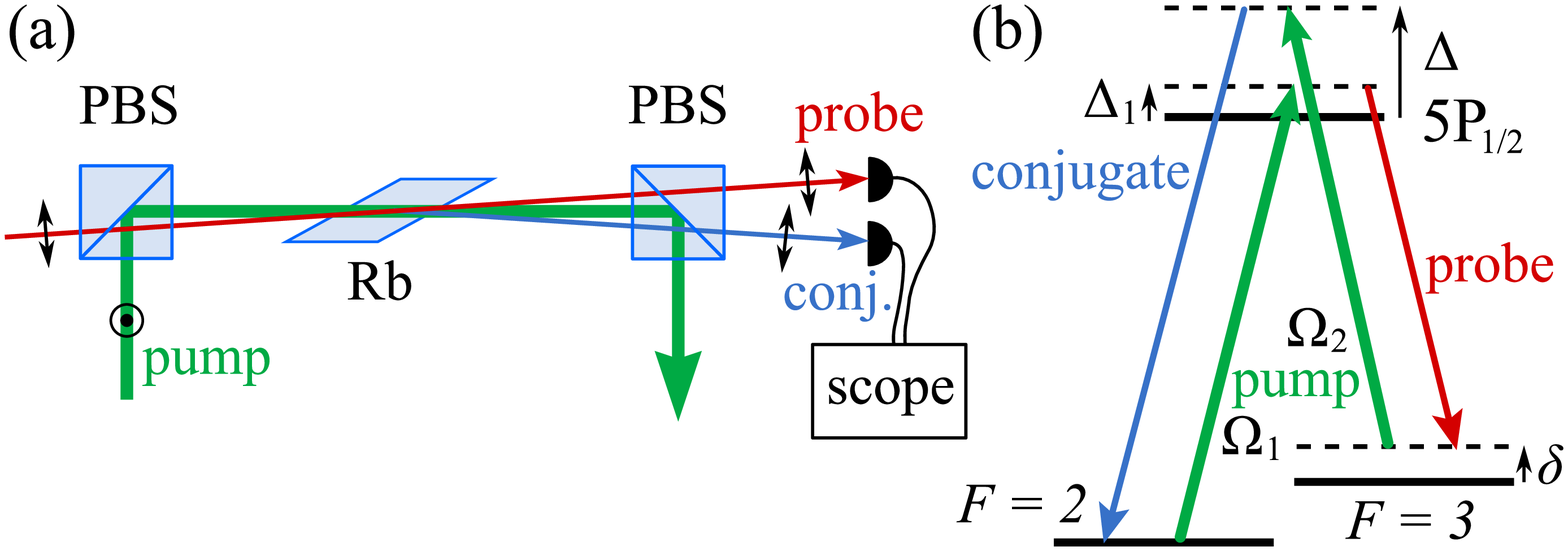}\\
	\includegraphics[height=.9\linewidth,angle=-90]{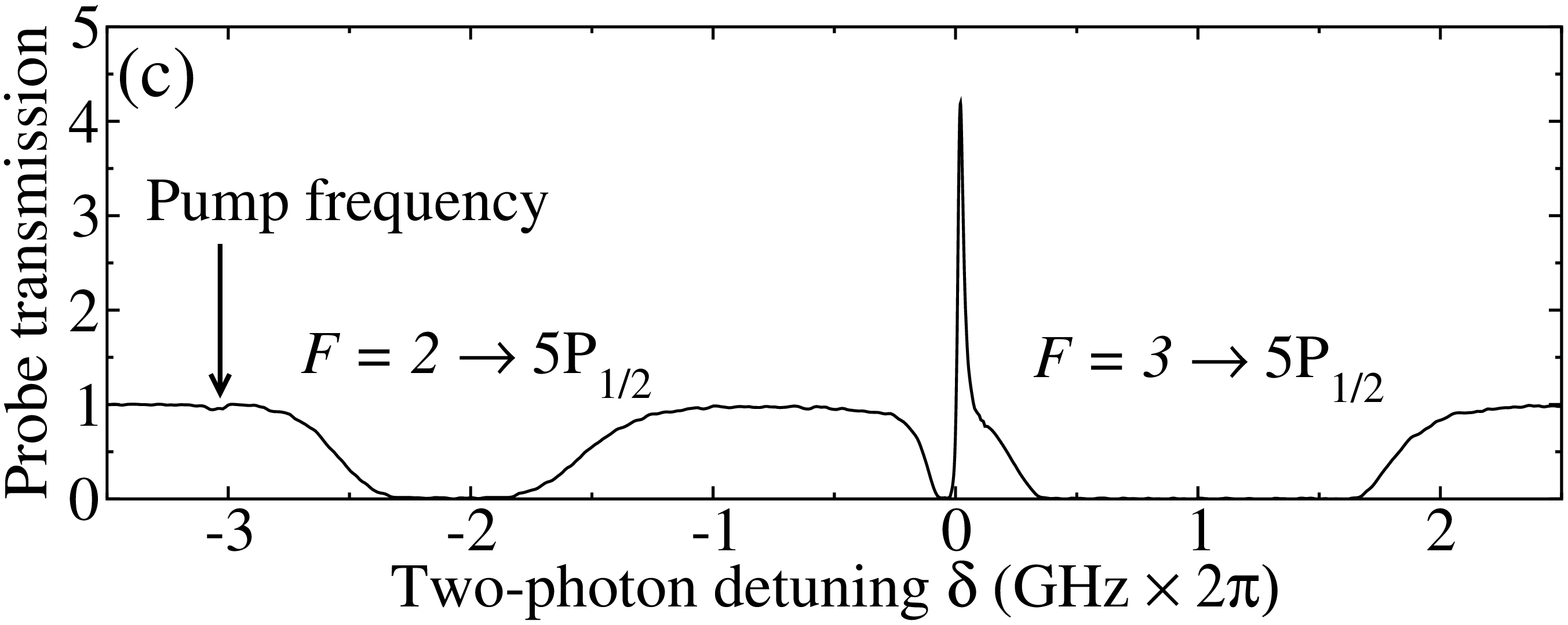}
    \end{center}
    \caption{(color online). (a) Experimental setup. PBS: polarizing beamsplitter. 
    (b) Energy-level diagram of the
    D1 line of $^{85}$Rb, showing the double-lambda scheme. Note that
    the pumps $\Omega_1$ and $\Omega_2$ are in fact the same laser beam. (c) Probe transmission
    profile versus two-photon detuning $\delta$.}
    \label{fig:exp}
\end{figure}

Our apparatus, which is essentially the same as the one described in
Ref.~\cite{mccormick2007}, consists of a linearly polarized,
continuous, strong (up to 280~mW) pump and a cross-polarized, pulsed, weak
(0.5 mW) probe propagating at a small angle ($0.5^\circ$) through a
2.5~cm-long $^{85}$Rb cell heated to $90^\circ$C-$140^\circ$C
(Fig.~\ref{fig:exp}a).  The pump and the probe are near resonant with
a Raman transition between the two hyperfine electronic ground states
of $^{85}$Rb, with a controllable detuning $\delta$
(Fig.~\ref{fig:exp}b), and have $1/e^2$ radii of 600~$\mu$m and
350~$\mu$m, respectively. The residual 2-photon Doppler broadening due
to the small angle is a few MHz. The detuning from the 5P$_{1/2}$
excited state is $\Delta_1 /2\pi \approx 850$~MHz, and the peak pump 
intensity (up to $45$~W/cm$^2$) is high enough to excite 
off-resonant Raman transitions with a detuning $\Delta / 2\pi\approx 4$~GHz
from the excited state.  The double-lambda~\cite{lukin1998} is closed
by the conjugate beam which emerges on the other side of the pump from
the probe, with the same polarization as the probe.  The combination
of the beam polarizations and the Zeeman substructure makes the system
a four-level system (the two virtual excited states are orthogonal).  
The probe
amplification is sharply resonant in $\delta$, as shown in
Fig.~\ref{fig:exp}c, with a gain that can reach 30.  The gain feature leads to a
strong dispersion of the index of refraction and thus a low group
velocity for the probe.

\begin{figure}[htb]
    \begin{center}
	\includegraphics[height=.9\linewidth,angle=-90]{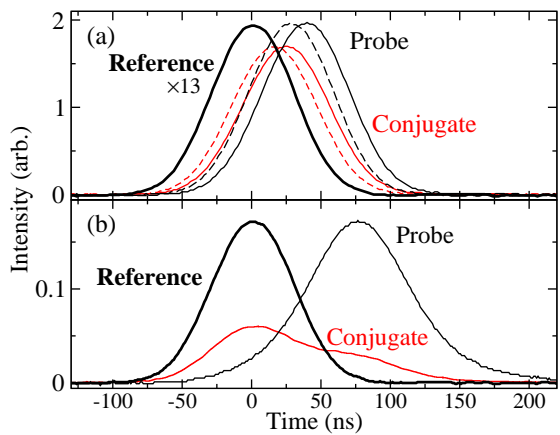}
    \end{center}
    \caption{(color online). (a) Slow-light effect near the peak 
    of the resonance.
    The reference pulse is magnified 13 times. 
    Thin black solid and dashed lines:
    probe pulses for the parameters ($\delta/2\pi$, pump power) equal to
    (10~MHz, 280~mW) and (22~MHz, 200~mW) respectively. 
    Color lines: corresponding
    conjugate pulses. The pulses (probe and matching conjugate) are 
    broadened by 5\% in the less
    retarded case, and 10\% in the more retarded case. (b) For a
    detuning $\Delta_1$ closer to the Raman absorption dip in
    Fig.~\ref{fig:exp}c, large delays and
    pulse breakups are observed (here for the conjugate).}
    \label{fig:slow}
\end{figure}

We measure the group velocity delay by recording the arrival time of a
70~ns-long (full width half maximum [FWHM]) gaussian probe pulse with
and without the atomic medium (reference pulse).
Figure~\ref{fig:slow}a shows an example in which the parameters are
set to provide a large  probe gain ($G=13$). By varying the two-photon detuning
$\delta$ and the pump intensity, one can tune the probe delay and
achieve a fractional delay larger than 0.5 such that the pulse remains
gaussian and is broadened by less than 10\% of its original width. 
This low level of
distortion is remarkable in comparison with that seen in some EIT
experiments~\cite{matsko2005}. The maximum delay corresponds to a
group velocity of $c/500$, where $c$ is the speed of light in vacuum.  
It is in general possible to tune the pump intensity, the pump
and probe detunings, and the temperature to achieve an overall gain of
unity.  Figure~\ref{fig:slow}a also shows the record of the conjugate
intensity. A striking feature is the emergence of the conjugate pulse
\emph{before} the probe pulse.  This relative delay is a fundamental property
of the dynamics of the system, and was predicted and observed in
Refs.~\cite{vanderwal2003,andre2005} in the case of resonance
on the ``lower'' lambda transition ($\Delta_1=0$).

Larger delays, shown in Fig.~\ref{fig:slow}b, can be achieved by setting $\delta$ between the gain peak
and the Raman absorption dip present at $\delta \lesssim 0$ (see
Fig.~\ref{fig:exp}c). 
The competition between large amplification and large absorption 
leads to some complex dynamics which can result in pulse
breakup, in a similar fashion to the dual-field solitons predicted to
exist in three-level systems~\cite{konopnicki1981}.
In this paper, we restrict ourselves to the regions of low absorption
where, in spite of complications associated with the use of hot atoms,
our system can be consistently described over a broad range of
parameters by the simple concepts developed in the theory of
Refs.~\cite{lukin2000,andre2005}.

We neglect the hyperfine splitting of the excited state. Averaged over
the Zeeman substructure, the dipole moments of all four transitions
are equal, which gives $\Omega_1$ and $\Omega_2$, the peak resonant
Rabi frequencies of the pump for the ``lower'' and the ``upper''
lambda respectively, the same value, denoted $\Omega$.  The
double-lambda system in its ideal incarnation operates in the limit
$\Delta_1 \ll \Delta$ and has the following crucial features.  First,
a ground state coherence is established by the ``lower'', more
resonant lambda. The coherence has a lifetime $1/\gamma_c$, limited by
magnetic fields, collisions and the transit time in the laser beams,
and corresponds to a dark state in which the absorption of the probe
is reduced (EIT). Second, the ``upper'', less resonant lambda slightly
perturbs this coherence and creates a resonant atomic polarization at
the probe and conjugate frequencies via 4WM, while keeping the
population in the excited state near zero.  The dynamics of the system
can thus be broken down into two intertwined processes: EIT and 4WM.

In the limit of a strong pump and low pump depletion, most of the
atomic population is in the ground state $F=3$, and the Fourier
components $\Ep(\omega)$ and $\Ec(-\omega)$ of the 
slowly-varying
envelopes of the probe and conjugate fields (of
wavevectors $\mathbf{k}_p$ and $\mathbf{k}_c$) obey the
equations~\cite{lukin2000,andre2005}:
\begin{eqnarray} 
    \label{prop1} 
    (i\omega + c\partial_z)\Ep &=&i
    \eta\Delta_R \Ec^* -\eta\left[i\left(\delta + \omega -
    {\textstyle\frac{\Omega^2}{4\Delta}}\right) +
    \gamma_c\right]\Ep\;\quad\\ 
    \label{prop2}
    (i\omega + c\partial_z)\Ec^* &=& -i \eta\Delta_R \Ep.
\end{eqnarray}
We assume perfect phase matching. In the limit of $\Omega^2/4\Delta_1
\gg \delta$, $\gamma_c$, and $\Delta \gg
\Delta_1$, $\gamma$
 (where $\gamma/2\pi = 6$~MHz is the linewidth of
the atomic transition), the coefficients in Eqs.~(\ref{prop1}) and
(\ref{prop2}) are~\cite{lukin2000} $\eta = g^2 N/[\Omega^2/4 +
\Delta_1(\delta + \omega + i\gamma_c)] \approx 4g^2 N/\Omega^2 = c/v_g \gg 1$
and
$\Delta_R=\Omega^2/4\Delta$. Here, $g^2 =
ck\wp^2/(2\varepsilon_0\hbar)$, $k = k_p \approx k_c$, $N$ is the
atomic density, and $\wp$ is the average dipole moment acting on the
probe and the conjugate.  

The interpretation of Eqs.~(\ref{prop1}) and
(\ref{prop2}) is straightforward. The probe field $\Ep$ is slowed down
by a factor $\eta$ due to the EIT interaction with the pump [second
drive term on the right-hand side of Eq.~(\ref{prop1})]. The EIT
resonance is light-shifted by the pump on the ``upper'' lambda and
occurs at $\tilde\delta \equiv \delta - \frac{\Omega^2}{4\Delta} = 0$.
In addition, $\Ep$ and $\Ec$ are cross-coupled with a coupling
constant $\alpha=\eta\Delta_R$, responsible for the 4WM amplification.
The presence of $\eta$ highlights the role of the longer interaction
time due to the slow-down effect in obtaining a sizeable nonlinear
coupling~\cite{lukin2001}. The other factor in $\alpha$, the so-called
Raman bandwidth $\Delta_R$, is the Rabi frequency of a fictitious
resonant Raman transition driven on both legs by the pump field and
with an intermediate Raman detuning $\Delta$.  As shown by the absence
of any dependence on $\Delta_1$, the 4WM dynamics is dominated by the
``upper'' lambda, which acts as a bottleneck.  The propagation
equations are asymmetrical.  The imaginary part of the coefficient of
the direct term for the conjugate [Eq.~(\ref{prop2})],
corresponding to a slow-light effect, is
negligible compared to the same term for the probe because $\Delta \gg
\Delta_1$\cite{lukin2000}.  The real part, corresponding to a 
Raman amplification,
scales as $1/\Delta^2$ and is negligible compared to the cross-term
$\alpha$ which scales as
$1/\Delta$~\cite{lukin2000,andre2005}. As a result, in the absence 
of the 4WM coupling (``bare''
fields), the probe and the conjugate propagate at velocities $v_g$ and
$c$, respectively.  The finite decoherence $\gamma_c$ translates into a
small absorption of the probe.  

Our system departs from the ideal case described by the expressions of
$\eta$ and $\alpha$ given above in many respects. Unlike the
experiments described in
Refs.~\cite{lukin2000,vanderwal2003,andre2005}, which were performed
with a resonant probe and a weak pump, our probe is tuned to the side
of the Doppler profile, in an already almost transparent region.  As a
result, the position of the gain peak is not tied as closely to a
narrow EIT window. It depends on the balance between the losses, which
include the Raman absorption dip and the absorption from the Doppler
broadened 1-photon transition, and the 4WM gain. Factors influencing
the peak position are the spread of values for $\Delta_1$ due to the
Doppler broadening, the spread of values for $\Omega_1$ and
$\Omega_2$ due to the Zeeman degeneracy, %
%
%
the contribution of the usual dispersion of the Doppler broadened
vapor to the slow-down of the
probe, and the only approximate phase matching. In practice, this
means that the position of the gain peak varies by up to 20 MHz
depending on parameters like the temperature, $\Delta_1$ and the probe
intensity.

In spite of the added complexity and the difficulty of directly
calculating $\eta$ and $\Delta_R$, we assume that the propagation
equations (\ref{prop1}) and (\ref{prop2}) are still valid over most of
the resonance peak, provided that the peak is at $\tilde\delta \approx 0$, 
and that $\gamma_c \ll 2\Delta_R$. 
Solving them in the limit
$\eta \gg 1$ and starting from a probe field $\Eo(\omega)$ and no
conjugate field, one finds:
\begin{eqnarray}
\label{sol1}
\Ep(\omega, z) & = & \Eo(\omega) 
\exp\left(i\sigma(\omega)\frac{z}{c}\right)\times\nonumber\\
& &\left[ \cosh \left(\xi(\omega)\frac{z}{c}\right) 
+ i\frac{\sigma(\omega)}{\xi(\omega)} \sinh
\left(\xi(\omega)\frac{z}{c}\right) \right]\\
\label{sol2}
\Ec^*(\omega, z) & = & \Eo(\omega) 
\exp\left(i\sigma(\omega)\frac{z}{c}\right)
\frac{\alpha(\omega)}{i\xi(\omega)} \sinh
\left(\xi(\omega)\frac{z}{c}\right)\quad
\end{eqnarray}
where $\xi(\omega) = \sqrt{\alpha(\omega)^2 -
\sigma(\omega)^2}$ and $\sigma(\omega) = 
\frac{\eta(\omega)}{2}(\tilde\delta + \omega + i
\gamma_c)$. Since $\gamma_c \ll 2\Delta_R$, $\xi(\omega)$
is real.  As expected, past an initial linear
growth of the conjugate, both fields grow exponentially in distance
with $(\xi- \frac{\eta}{2}\gamma_c)/c$ as the linear gain
coefficient. In the limit $\tilde\delta = 0$, one has
$\xi \approx \alpha$.

The solutions also contain information about the propagation delay.
Equations (\ref{sol1}) and (\ref{sol2}) show that
the fields accumulate a phase
across the medium, denoted $\theta(\omega)$. 
The main contribution to the group
delay $\frac{\mathrm{d}}{\mathrm{d}\omega}\theta(\omega)$ for both
fields comes from the first exponential and gives a common delay $\tau
= \left.\frac{\mathrm{d}}{\mathrm{d}\omega}\Re[\sigma(\omega)]
z/c\right|_{\tilde\delta = 0,\omega = 0} = \eta z/2c$.  In other words,
the probe and the conjugate are slowed down by half the ``bare''
slow-down factor $\eta$. The probe experiences an extra delay
due to the second term in Eq.~(\ref{sol1}).  At large gain, the cosh
and sinh functions are equal and the additional delay is $\Delta\tau =
\left.  \frac{\mathrm{d}}{\mathrm{d}\omega}\Re[\sigma(\omega) /
\xi(\omega)] / (1 - \Im[\sigma(\omega) / \xi(\omega)])
\right|_{\tilde\delta = 0,\omega=0} = \eta/[2\xi -
\eta\gamma_c] \approx \eta/2\xi$.  At low
gain, a first order expansion in $z$ gives $\Delta \tau = \eta
z/ 2c$. The picture emerging from this analysis is the following: the
conjugate pulse is created without delay by the probe and travels
at a velocity $2v_g$ ($\ll c$). The probe pulse travels initially at a
velocity $v_g$ and then locks onto the conjugate by accelerating to
$2v_g$ when the delay reaches $\eta/2\xi$ (at a gain
close to 2).

\begin{figure}[htb]
    \begin{center}
	\includegraphics[height=.9\linewidth,angle=-90]{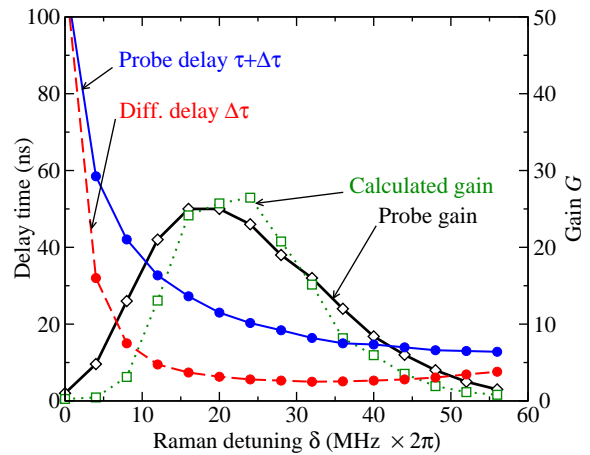}
    \end{center}
    \caption{(color online). Two-photon detuning scan at a temperature
    of $140^\circ$C, using 120~ns-long pulses.  The bare state
    2-photon resonance corresponds to $\delta = 0$.  The calculated
    gain, inferred
    from the delays, assumes linear losses equal to 14\% of the peak
    linear gain.}
    \label{fig:delays}
\end{figure}

We test this interpretation by first scanning the two-photon detuning
$\delta$, using 120~ns-long (FWHM) probe pulses and a pump power
of 280~mW, which corresponds to a spread of $\omega$ of 10~MHz
around zero and to $\Omega/2\pi = 420$~MHz.
Figure~\ref{fig:delays} shows the measured gain, the probe delay $\tau
+ \Delta\tau$, which includes the contribution of all the retardation
effects, and the differential delay $\Delta\tau$.  The contribution to
the probe delay of the usual dispersion effect,  which is measured
with the pump intensity strongly reduced, is found to be $8\pm2$~ns.
From the experimental data and the theoretical expressions of $\tau$
and $\Delta\tau$, one can deduce a value of $\eta$ and $\xi$ for each
$\delta$, in the limit of large gain and small $\gamma_c$.  Inserting
these values into Eq.~(\ref{sol1}) (with $\tilde\delta \approx 0$) and
adjusting $\gamma_c$ to $0.5\gamma$ (making the linear losses equal 
to $14\%$ of the 
peak linear gain)~%
\footnote{This value of $\gamma_c$ is sensitive to the timing errors
($\pm1$~ns) and is only an indication of the order of magnitude.
For $G=3$, it corresponds to 15\% of absorption, 
which is compatible with the level of squeezing
measured in Ref.~\cite{mccormick2007}.}  
, one can reproduce the gain curve with reasonable accuracy.  For our
parameters, the EIT resonance is light-shifted to $\delta =
11$~MHz$\times 2\pi \approx 2\gamma$, close to the observed gain
maximum ($\tilde\delta = 6$~MHz$\times 2\pi \approx \gamma$).  It can be
shown that in the above calculations, the approximation $\tilde\delta
\approx 0$ is valid as long as $\tilde\delta \ll 2\Delta_R$
($4\gamma$ for our beam parameters).
For $|\tilde\delta|$ larger than a few $\gamma$, that is to say in the
wings of the gain peak, the approximation is
expected to break down.  

\begin{figure}[htb]
    \begin{center}
	\includegraphics[height=.9\linewidth,angle=-90]{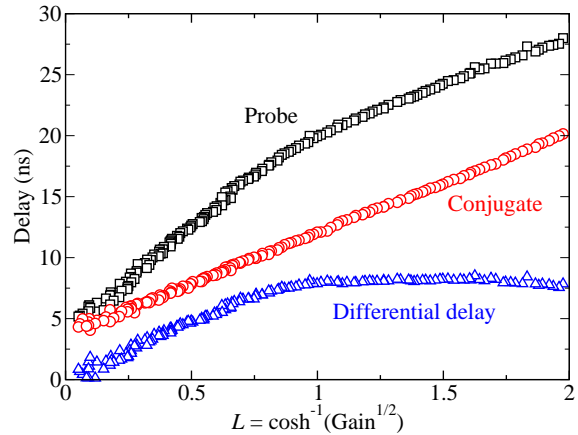}
    \end{center}
    \caption{(color online). Probe delay $\tau + \Delta\tau$, conjugate 
    delay $\tau$, and differential delay $\Delta\tau$ as a function of 
    a pseudo propagation distance.}
    \label{fig:locking}
\end{figure}

Next, we directly observe the locking between the probe
and the conjugate during the propagation.  It is impractical to
continuously vary the distance of propagation and we instead vary the
atomic density $N$ via the temperature, which is equivalent.  
Indeed, $\sigma$ and
$\xi$ are proportional to $N$ through their dependence on
$\eta$, and changing $N$ is like renormalizing $z$ in the solutions
(\ref{sol1}) and (\ref{sol2}). According to (\ref{sol1}), the
renormalized propagation length $L$ is related to the probe intensity
gain $G$ by $L= \cosh^{-1}(\sqrt{G})$.
The detuning $\delta$ is set to 15 MHz$\times 2\pi$, near the gain 
maximum, the pump power
is still set to 280~mW, and the measured
delays as a function of the renormalized distance for a temperature
scan of $50^\circ$C around $120^\circ$C are shown in
Fig.~\ref{fig:locking}. The two regimes of
propagation are very clear. First, the pulses separate in time, and second they lock
to each other at a fixed delay. The time offset at the origin is not
well understood.
A direct evaluation of
$2\Delta/\Omega^2 \approx \Delta\tau$ in the ideal case using
our beam parameters gives a value of 7~ns, comparable to the one measured
near the gain maximum (see Fig.~\ref{fig:delays}). 
We checked qualitatively that $\Delta\tau$ increases when the pump
intensity decreases.  It is worth noting that the detail of the
low-gain transient regime depends on the initial conditions. For
instance, swaping the frequencies of the probe and the conjugate would lead
to an initial propagation in which the probe travels at a velocity $c$
while the conjugate travels at a velocity $2v_g$.

Finally, an important feature of the model is that the gain saturates
with the pump intensity.  For a gain peak location $\delta$ and a
decoherence $\gamma_c$ of the order of $\gamma$, $\xi$ saturates  when
$\Omega \gg 2\sqrt{\Delta\gamma} = 300$~MHz $\times 2\pi$. In agreement
with this prediction, we observe that $G$ starts saturating at our
operating intensity.  

To conclude, we have observed the ultraslow propagation of probe and
conjugate pulses with matched shapes and group velocities through a rubidium
vapor. The study of the coupled propagation gives access to the atomic
dynamics through a simple model that reflects a few key concepts.
Although the hypothesis of the model does not match precisely the
conditions of the experiment, our findings on slow propagation and
delay locking of the probe and conjugate pulses are generic to the
double-lambda system.  The quality of the retardation effect in terms
of fractional delay and absence of loss and distortion suggests the
possible existence of a dual-field soliton, which would be the result
of higher order terms in the propagation equations. Such a soliton has
been predicted in related 4WM schemes~\cite{deng2005}.

It should also be noted that this double-lambda scheme is known 
to generate relative-number
squeezed twin beams~\cite{mccormick2007} (as well as correlated
photons~\cite{kolchin2006}). Extending this
semi-classical pulse
theory to the quantum correlated beams raises two comments. First, as
pointed out in Ref.~\cite{andre2005}, the time lag between the probe
and the conjugate should be the limiting factor to the squeezing
bandwidth observed in Ref.~\cite{mccormick2007}. Second, the system
could be used with gain close to unity to slow light in the quantum
regime, while preserving nonclassical correlations,
possibly more efficiently than with EIT alone~\cite{akamatsu2006}.

CFM was supported by an IC Postdoctoral Fellowship.
We acknowledge very helpful discussions with A. M. Marino.

\bibliography{slow}

\end{document}